\RequirePackage[hyphens]{url}
\documentclass[onecolumn,amsmath,amssymb,groupedaddress,superscriptaddress,11pt]{revtex4-2}
\usepackage{times}
\usepackage[hyphens]{url}
\usepackage[utf8]{inputenc}
\usepackage[T1]{fontenc}
\usepackage{graphicx}         % For PDF figures
\usepackage{xcolor}
\definecolor{color1}{HTML}{69a1b0}
\definecolor{color2}{HTML}{8fcece}
\definecolor{color3}{HTML}{ef476f}

\begin{document}
\title{\Large{Quantum key distribution for data center security -- a feasibility study}}

% Nitin Jain, DTU Physics\\
% Tobias Gehring, DTU Physics\\
% Ulrich Hoff, DTU Physics\\
% Jesper Rodenberg, Danske Bank\\
% Marco Gambetta, KPMG Denmark\\

\author{Nitin Jain}
 \email{nitin.jain@iitbombay.org}
 \affiliation{%
 Center for Macroscopic Quantum States (bigQ), Department of Physics, Technical University of Denmark, 2800 Kongens Lyngby, Denmark}
 \author{Ulrich Hoff}
\affiliation{%
 Center for Macroscopic Quantum States (bigQ), Department of Physics, Technical University of Denmark, 2800 Kongens Lyngby, Denmark}%
 \author{Marco Gambetta} 
 \affiliation{KPMG P/S Dampfærgevej 28, 2100 København, Denmark }
\author{Jesper Rodenberg}%
\affiliation{%
 Danske Bank, Edwin Rahrs Vej 40, 8220 Brabrand,  Denmark}%
\author{Tobias Gehring}
\email{tobias.gehring@fysik.dtu.dk}
\affiliation{%
 Center for Macroscopic Quantum States (bigQ), Department of Physics, Technical University of Denmark, 2800 Kongens Lyngby, Denmark}%

\date{\today}% It is always \today, today,

\begin{abstract}
Data centers are nowadays referred to as the digital world's cornerstone. Quantum key distribution (QKD) is a method that solves the problem of distributing cryptographic keys between two entities, with the security rooted in the laws of quantum physics. This document provides an assessment of the need and opportunity for ushering QKD in data centers. Together with technical examples and inputs on how QKD has and could be integrated into data-center like environments, the document also discusses the creation of value through future-proof data security as well as the market potential that QKD brings on the table through e.g., crypto-agility. While primarily addressed to data center owners/operators, the document also offers a knowledge base to QKD vendors planning to diversify to the data center market segment. 
\end{abstract}

\maketitle

\newpage
%\tableofcontents
\clearpage

\section{Motivation}\label{sec:mot}
Data and information protection is a central requirement penetrating society from the level of individual citizens, to industries and businesses, and all the way up to national states, and it is a topic of ever growing importance. Protecting data and information can be achieved either by restricting physical access, e.g., by using vaults and safes, or by obscuring the information to the unauthorised entity by cryptographic methods. Often a combination of both is used. In either case, the security robustness is closely connected with technological development. 

At a very basic level, encryption of data is achieved by admixing randomness to the initial plaintext information, resulting in a cryptotext that reveals as little information as possible about the plaintext. The randomness is provided in the form of an encryption key and only with a knowledge of the key can information be restored. Over time, the technological implementation of cryptographic schemes has developed from rudimentary manual devices, encoding one or the other variation of a substitution cipher, to intricate mechanical and electrical devices, and to the quite-recent algorithm-based systems. 
\begin{figure}[!hbpt]
\centering
\includegraphics[width=0.99\linewidth]{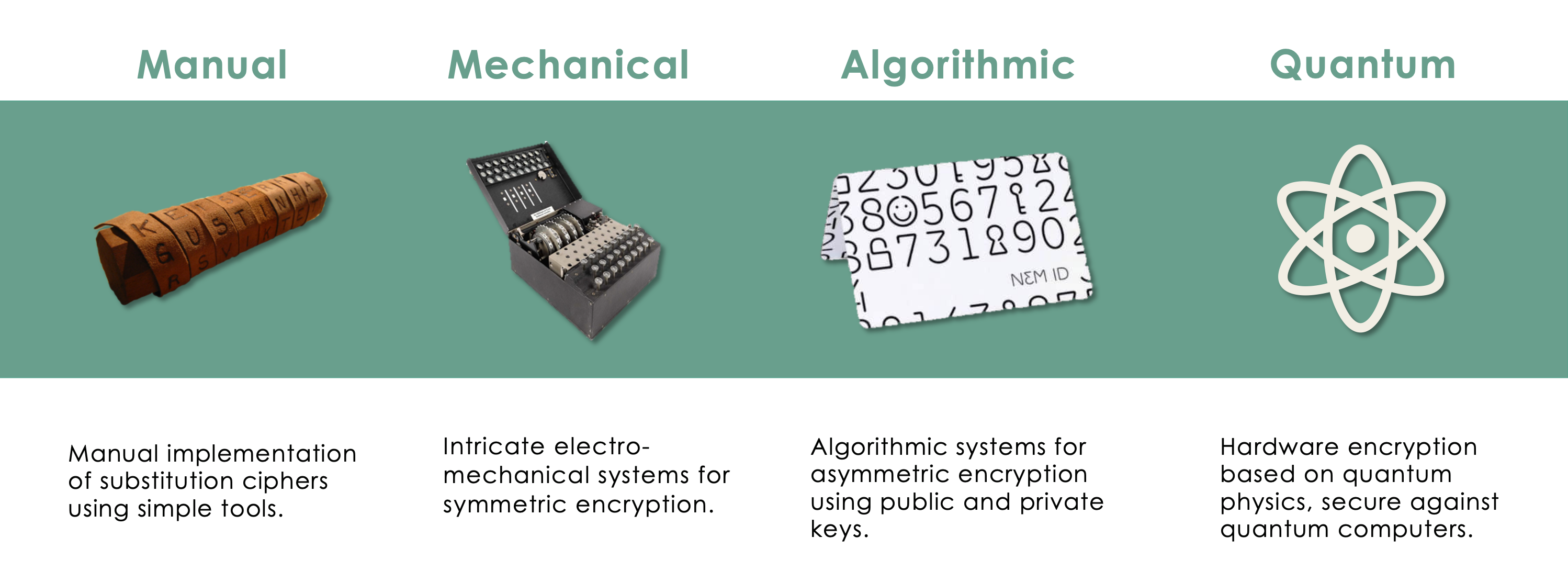}
\label{fig:evo-crypto}
\end{figure}
Indeed, the modern digital world has become highly dependent on cryptography, in particular public key cryptography, and the majority of currently employed schemes are based on the assumed complexity of specific mathematical problems, such as factorisation of large integers to their prime number components. However, with the advent of quantum technology, and in particular quantum computing, the security provided by such schemes will be rendered inadequate. Thus, development and implementation of quantum-safe capabilities is critical for maintaining data security and integrity for critical applications and infrastructure.

Quantum key distribution (QKD) harnesses quantum physics for the task of distributing keys securely between two parties. Amongst all emerging technologies that leverage quantum effects to outperform conventional techniques in terms of functionality, speed, and sensitivity, QKD is arguably the most mature. Recent exploits (within Europe) include deployment of the first quantum network across the borders of Italy, Slovenia, and Croatia~\cite{qkdnews_ItSaCa}, commercially available QKD metro network in the UK~\cite{qkdnews_UK}, and establishment of an intercity QKD infrastructure in Poland~\cite{qkdnews_Pol}. QKD has the potential to deliver unprecedented data protection, even against attacks from quantum computers. Consequently, it is foreseen that QKD shall play an important role in future high-security data infrastructures.  
\begin{figure}[!t]
\centering
\includegraphics[width=0.85\textwidth]{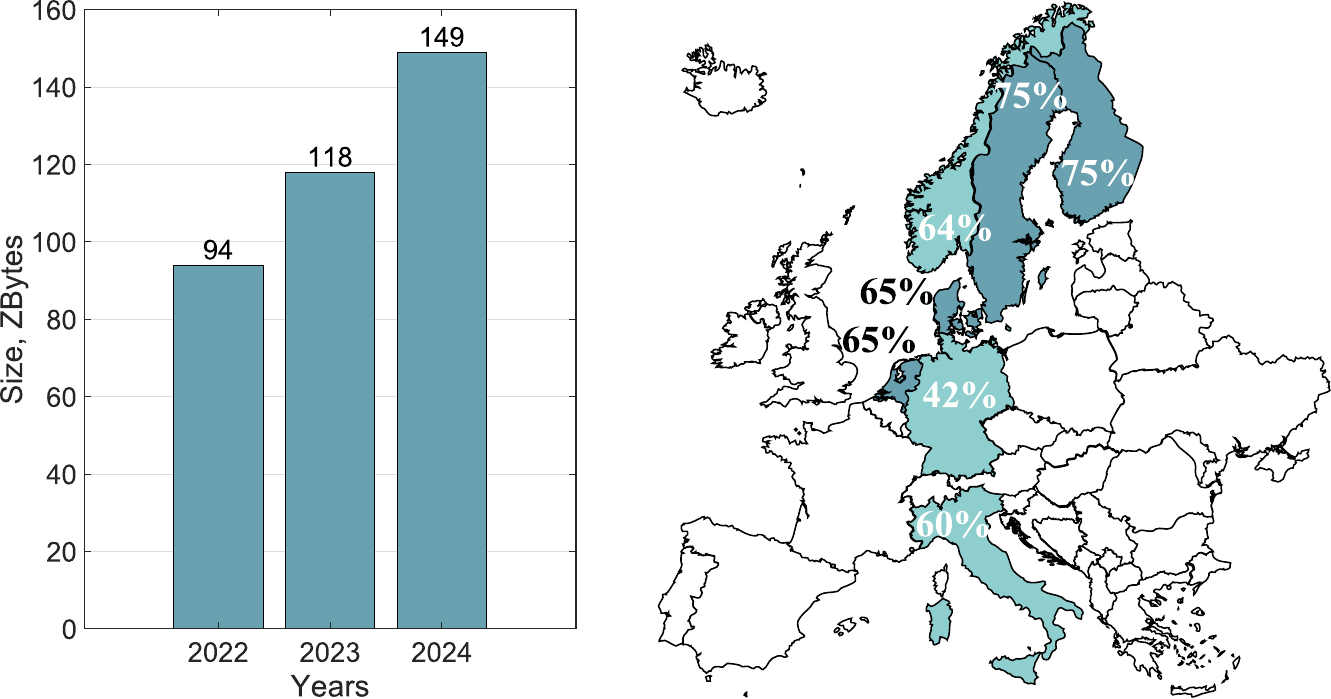}
\caption{{(Left) Global data volume forecast~\cite{data_vol_growth}. One zetabyte (ZByte) = $10^{21}$ bytes. (Right) Share of companies using cloud services in various European states, as of 2021~\cite{Hintemann2021}.}}
\label{fig:evo-globData_cloudcovEU}
\end{figure}

In 2022, humanity created, copied, and consumed an estimated 94 zetabytes of data~\cite{data_vol_growth}, and the trend is only going towards larger data volumes; see Fig.~\ref{fig:evo-globData_cloudcovEU}(left). However, this rapid acceleration has also brought in many challenges, with a primary one being an increased incidence of cyberattacks. For instance, the number of distributed denial of service (DDoS) attacks has grown with a compounded annual growth rate of 14\% since 2018, with $>$15 million DDoS attacks projected in 2023~\cite{cisco_annual}. Data breaches involving public and private enterprises now regularly make it to mainstream news. Ensuring cybersecurity is thus a top priority for a multitude of organizations, ranging from businesses and governments to hospitals and universities. 

Hand in hand with the exploding global data consumption, data centers play an increasingly importance role for society. To keep up with the demand, the number of data centers has skyrocketed. As of January 2021, the number of data centers in the top-10 hosting countries amounted to 5626, and over 515 facilities were further built or expanded in the same year. The global revenue from the colocation data center market amounted to about USD 50.5 billion in 2021, and the industry revenues are expected to increase to over USD 136 billion by 2028~\cite{data_center_growth}. 

The main driver for this growth are a rapid adoption of cloud computing in particular and digital technologies in general. A recent report from the Borderstep institute corroborates this trend~\cite{Hintemann2021}. The right side of Fig.~\ref{fig:evo-globData_cloudcovEU} (using data from this report) provides a glimpse into the cloud services' usage by companies located in various European countries and having 10 or more employees. In the Nordic region, cloud computing is clearly a dominant player, with Sweden (75\%), Finland (75\%), Denmark (65\%) and Norway (64\%), being well above the European average level. While Germany (42\%) seems to score lower compared to Holland (65\%) and Italy (60\%), cloud adoption is nonetheless increasing at a breathtaking pace: Between 2018 and 2021, the number of cloud-service-using companies in Germany jumped up by a factor of two~\cite{Hintemann2021}. 
%%%%%%%%%%%%%%%%%%%%%%%%%
\section{Background}
%%%%%%%%%%%%%%%%%%%%%%%%%
\subsection{Cybersecurity \& OSI model}
The Open Systems Interconnection (OSI) model is a reference model that describes the flow of information between devices on a network. For instance, an HTTP request made by a browser on one device to a web server hosted on another device involves data moving through several abstraction layers. Distinct communication protocols enable interaction on each of the seven different layers: Physical, Data Link, Network, Transport, Session, Presentation, and Application (starting from the lower to upper) of the OSI model. From a cybersecurity perspective, these protocols may also aim to provide confidentiality, authentication and integrity throughout the journey between the layers. 
% The Open Systems Interconnection model (OSI model) is a reference model that describes how information from an application in one computer on a network moves through a physical medium to a software application on another computer. The model is divided into seven different abstraction layers (starting from the lower to upper layers): Physical, Data Link, Network, Transport, Session, Presentation, and Application.
% For each layer there are distinct communication protocols that enables interaction with the corresponding layers. The protocols aim to provide confidentiality, authentication and integrity throughout the journey between layers and in a network from computer to computer. The OSI model is an older reference model, but is still widely used as it helps with the visualization and commination of how networks operate to isolate and troubleshoot problems. A simpler TCP/IP model is more akin to how the modern internet operates, and it encompasses many of the same elements as the OSI model, including communication protocols. However, most of the communication protocols in both models, rely on classical algorithms based on mathematical complexity of the integer factorization problem, which becomes vulnerable to quantum computing.

One way to ensure data confidentiality, i.e.\ to prevent sensitive information from unauthorized access, is to ``lock'' it with a secret key --- secretly chosen string of bits --- known only to the authorized parties. The most well-known cryptographic method to lock / unlock information is symmetric encryption / decryption. In the OSI model context, an encryption algorithm employed on say the Data Link layer uses a secret key to encrypt the data coming down from the Network layer, before getting transmitted on the communication channel, i.e., the Physical layer. An authorized receiver connected to the channel uses the same secret key to decrypt and obtain the original message on the Data Link layer, and then passes it upwards onto the Network layer. 
% In a communication channel context, the sender encrypts data using an encryption algorithm and a secret key, transmits it on the channel that connects the receiver, who uses a decryption algorithm and the same key to obtain the original message. 
An adversary, i.e., an unauthorized and malicious entity, could access the channel and copy the encrypted data in transit but (at least theoretically) cannot decrypt it without having the key. 
%%%%%%%%%%%%%%%%%%%%%%%%%%%%%%%%%%%%%%%%%%%%
\subsubsection{Computational (in)security and the problem of key distribution}\label{bg:prob_key_distb}
In practice however, a powerful adversary could use computational resources to break confidentiality. This could involve taking advantage of design flaws in the encryption algorithm, e.g., information leakage that eases the task of cryptanalysis, poorly chosen security margins, e.g., short length keys that can be exhaustively searched for, etc. Contrarily, an encryption scheme is \emph{computationally secure} if it is able to withstand attacks and hold confidentiality over a reasonably long period.  
\begin{figure}[!t]
\centering
\includegraphics[width=0.95\linewidth]{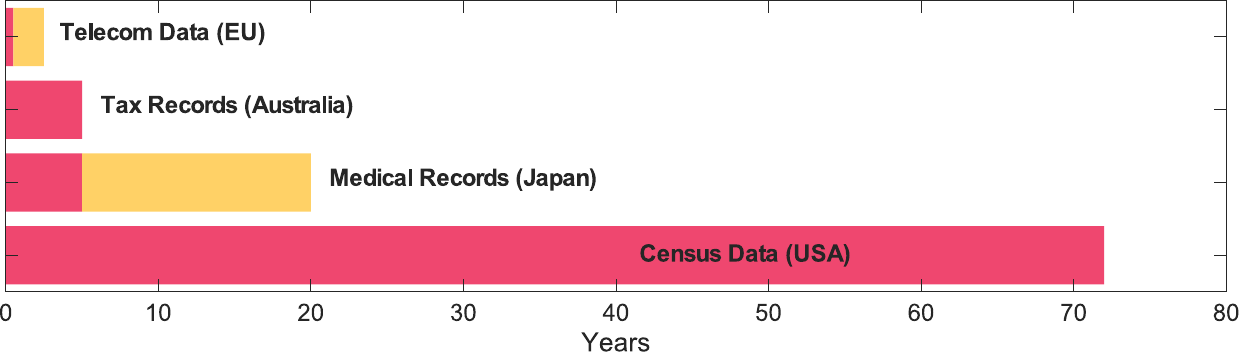}
\caption{Retention requirements for different types of data / records in different parts of the world (Sources: \cite{data_retpol_EU, tax_rec_aus, nakamura_current_2006, census_USA}). The red bars depict mandatory periods while additional recommended periods (on top of the mandatory ones) are shown by yellow bars. Data retention policies generally focus on storage, maintenance, and retrieval of data, however, depending on how sensitive the data is by nature, may require information security as another critical parameter. For instance, the Census Bureau in USA is legally required to protect the collected personal information and keep it \emph{confidential} for 72 years~\cite{census_USA}.}
\label{fig:data_ret}
\end{figure}
Depending on the underlying nature of data, see Fig.~\ref{fig:data_ret} for some examples, some algorithms may clearly be unable to guarantee confidentiality for the designated period.

The current workhorse of symmetric key algorithms is the so-called advanced encryption standard (AES) that typically uses either 128-bit or 256-bit long keys. The key lengths are sufficiently large to rule out any exhaustive search even over several millennia, and a focused cryptanalysis research has vetted design flaws in the algorithm itself. In fact, AES is now the most widely used symmetric key algorithm for securing data on the Internet, and is estimated to have had an economic impact of USD 250 billion in the first 21 years of its existence~\cite{leech_economic_2018}. 

In general, using well-vetted and regularly updated cryptographic algorithms reduces the risk of cybersecurity breaches to a minimum. Furthermore, by employing a good quality source of random numbers to construct sufficiently long keys as well as periodically changing them, one can mitigate the threat that an adversary can easily guess the key. Nonetheless, the \emph{task of distributing the keys to authorized parties prior to encryption remains}. 

An obvious and simple solution is that of using pre-shared keys and/or a trusted courier to distribute the keys. In the former, the encryption devices are loaded with a large amount of the same random bits at the factory. After deployment, these bits can be used as the key material (after a proper synchronization) for encryption and decryption of messages. The latter solution involves a courier or messenger traveling to the deployment sites for providing (or renewing) the secret key material. Apart from being limited in scope and potentially unviable from an economic perspective, such key-sharing solutions involve an element of trust that cannot be analysed in a cryptographic framework. 

The most popular and widespread method as of today to solve this task uses asymmetric cryptographic techniques~\cite{Diffie1976,Rivest1978}. Each authorized party utilizes a pair of public and private keys. As the names suggest, the public key is openly accessible to anyone, while the private key is exclusive. Two parties can use this public-private key combination to enable secure computation of symmetric keys (which may be used later for say AES encryption). A notable example are so-called session keys used in the transport layer security (TLS) protocol. Such keys are discarded after a `session', a preset time-interval, and fresh ones are generated. Depending on the variants of the asymmetric cryptographic algorithms used for implementing TLS in practice, a premaster secret or a Diffie-Hellman parameter, securely exchanged by the two parties using public and private keys, is used for computation of the session keys~\cite{tls_handshake}. 
% This is one reason behind the usage of session keys, i.e., discarding after a preset time interval, or a session,  and regenerating fresh
%To distribute for instance a so-called session key (to be used later say for AES encryption), one of the authorized parties obtains the public key belonging to another, encrypts the session key with this public key, and sends it on the communication channel to the other party. The construction of the algorithm using so-called ``one-way functions'' enables the other party then to decrypt that session key using their private key. 

Compared to symmetric encryption schemes, public key algorithms are \emph{computationally insecure}. With enough computing power, the public-private key pair can be obtained (even if inefficiently). Furthermore, discovery of new mathematical methods and technological advances in conventional computing can make that process efficient. A major threat actually comes from quantum computing, where efficient methods to break the security of public key algorithms are already known~\cite{Shor1994}. 
%%%%%%%%%%%%%%%%%%%%%%%%%%%%%%%%%%%%%%%%%%%%
\section{Quantum key distribution (QKD)}
%%%%%%%%%%%%%%%%%%%%%%%%%%%%%%%%%%%%%%%%%%%%
The physics behind quantum computing (and thereby, also the threat to the public key cryptographic infrastructure) remarkably also provides a solution to the problem of key distribution across an open communication channel. By harnessing quantum phenomena such as superposition and uncertainty, the authorized parties can share quantum correlations that can be converted to a common bit-string. 

Notably, an adversary who attempts to obtain that common bit-string by eavesdropping on the communication degrades the correlations. The authorized parties can not only detect but even quantify these actions of the eavesdropper. Thereafter, there are two possible scenarios depending on the degree of eavesdropping: the parties either distill from their common bit-strings a shorter but more secure key for encryption or an eavesdropping alarm is raised and the encryption process is aborted. In both scenarios, the confidentiality of the data remains intact. 

Such a quantum key distribution (QKD) method is thus \emph{information-theoretically secure}~\cite{bloch_overview_2021}, and coupled with one time pad (OTP) encryption~\cite{vernam19, Shannon1949}, it guarantees that only a computationally infeasible brute-force attack can break the confidentiality. OTP however requires that the length of the key should be at least equal to the length of the data, and that the keys should always be `fresh', i.e., never repeated. 

\vspace{2ex}
\fbox{
%\colorbox{color2}{
\begin{minipage}{0.92\textwidth}
\textit{\large Foundational principles on which QKD is based}
\begin{itemize}
    \item[$\blacktriangle$] \textbf{Quantization \& indivisibility of quanta}: The concept of quantization and indivisibility of quanta~\cite{Planck1901,Gerlach1922,Dirac1930} provides us with photons that can carry quantum information across the communication channel. The sender and receiver typically encode and decode, respectively, information in some degree of freedom of each photon. 
    \item[$\blacktriangle$] \textbf{Heisenberg uncertainty principle}: A fundamental indeterminacy in being able to simultaneously measure a pair of properties, e.g., position and momentum, of the photon~\cite{Heisenberg1927} is crucial for limiting what an adversary can know without introducing perturbations. 
    \item[$\blacktriangle$] \textbf{No-cloning principle}: The principle prohibits the adversary from creating perfect clones of a single photon without knowing all of its original properties before~\cite{Wootters1982}. Otherwise, the adversary could simply clone and store copies of the photons leaving the sender, and later, perform the same measurements as the receiver. 
\end{itemize}
\end{minipage}
}
%%%%%%%%%%%%%%%%%%%%%%%%%%%%%%%%%%%%%%%%%%%%
\subsection{Basic principles}
%%%%%%%%%%%%%%%%%%%%%%%%%%%%%%%%%%%%%%%%%%%%
At its core, any physical QKD system is a communication system that additionally exploits quantum physics to facilitate a distribution of quantum correlations between the authorized parties. The actions of the eavesdropper also result in correlations with the sender and/or receiver. With the help of a theoretical security model, the authorized parties can quantify all these correlations and determine if they can eventually obtain a secret key; otherwise, they declare the channel to be unsafe for communication. 

Photons are the only quantum-physical objects capable of providing both of these communication and cryptographic capabilities. In classical telecommunication networks, photons are routinely used as carriers of information. The digital bit information is coded onto one of many degrees of freedom or `variables' of the photon using optical components such as modulators and detectors. The main difference between classical and quantum communication is in the number of photons used to convey say one `symbol' of information: the former uses several orders of magnitude larger photon number per symbol. 

Using one or just a few photons is in fact essential to the security guarantee provided by QKD, due to the no-cloning principle. Finally, the choice of the variables to be used for coding information is guided by the uncertainty principle to ensure that any eavesdropping activity necessarily results in perturbations that degrade the quantum correlations. 
%%%%%%%%%%%%%%%%%%%%%%%%%%%%%%%%%%%%%%%%%%%%
\subsection{Implementation}
%%%%%%%%%%%%%%%%%%%%%%%%%%%%%%%%%%%%%%%%%%%%
A QKD system performs a sequence of steps called a ``QKD protocol'', which is operated over a pair of channels. The \emph{quantum channel} for relaying photons---encoded by the sender and decoded by the receiver---is implemented using a fiber-optic or an atmospheric line-of-sight link. 

After this quantum stage of the QKD protocol, the classical (processing) stage begins. The authorized parties apply a series of mathematical transformations on their respective (en/de)coded symbols and exchange messages on an \emph{authentication channel}, in order to estimate the parameters of the quantum channel and quantify the actions of the eavesdropper. The authentication channel can be a regular telecommunication channel: its purpose is to ensure that the authorized parties indeed communicate with each other. 
\begin{figure}[!hp]
\centering
\includegraphics[width=1.0\linewidth]{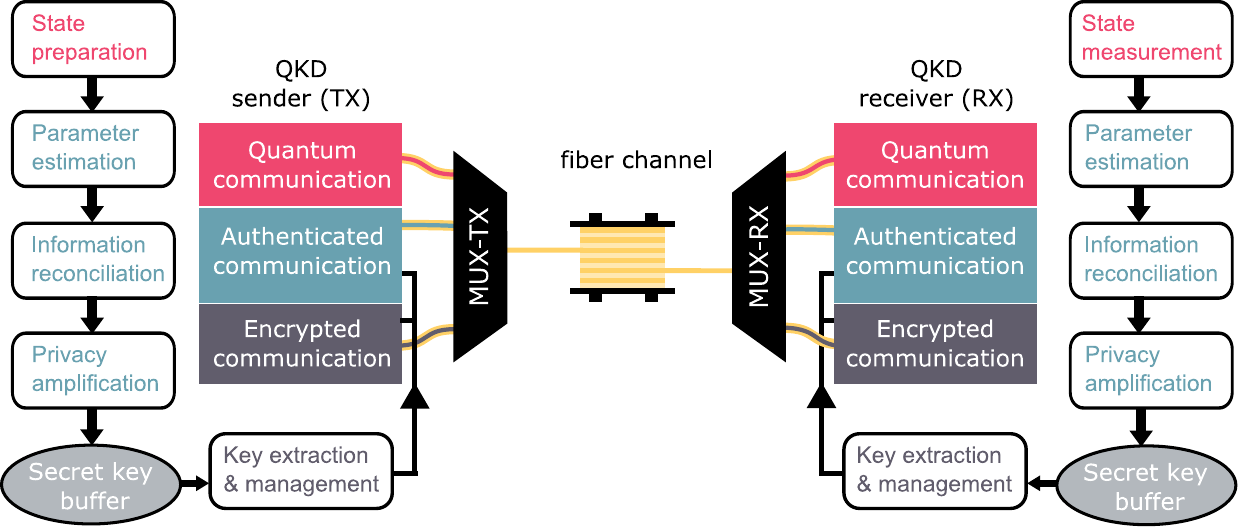}
\caption{Point-to-point link encryption using QKD. Quantum communication modules at the sender and receiver distribute quantum correlations over the fiber channel through preparation and measurement of quantum states of light. The ensuing classical phase of the QKD protocol then attempts to distill secret bitstreams from these correlations, and whenever successful, deposits them to secret key buffers on each side. The modules responsible for the authenticated and encrypted communication are assumed to have suitable optical front-ends in order to use the same fiber channel. Typically the quantum communication is unidirectional, whereas the authenticated and encrypted communication is bidirectional. MUX: Multiplexer, TX: Transmitter, RX: Receiver.}
\label{fig:scheme}
\end{figure}

At the end of a successful QKD protocol, both the sender and receiver possess the (same) secret key, on which the eavesdropper's knowledge has also been reduced to an insignificant level. This secret key can be used for symmetric encryption and decryption on a \emph{communication channel}. 

It is possible to implement all of the aforementioned channels on the same physical medium using multiplexing techniques. For instance, as illustrated in Fig.~\ref{fig:scheme}, three different optical wavelengths could be used for supporting the quantum, authentication, and encrypted communication traffic on a single fiber-optic link connecting the sender and receiver. A portion of the key generated at the end of the QKD protocol is used for the purpose of maintaining the authenticated channel. 

Depending on the types of variables used for coding information, there are two main categories of QKD implementations: discrete and continuous. Discrete-variable (DV) QKD implementations rely on encoding using a discrete set of degrees of freedom at the sender and single-photon detectors for decoding at the receiver. Polarization or (relative) phase of photons are the best known examples of variables employed by DVQKD implementations. In continuous-variable (CV) QKD implementations, the information is encoded by the sender in continuously valued variables such as amplitude and phase. The receiver decodes this information using coherent detection facilitated by a so-called local oscillator. 

Given some channel parameters, the main output of interest from a QKD protocol is the amount of the secret key it is able to generate. One typically uses a secret key rate (SKR), i.e., the number of secret key bits generated in a given period of time. 
\begin{figure}[!thp]
\includegraphics[width=0.67\textwidth]{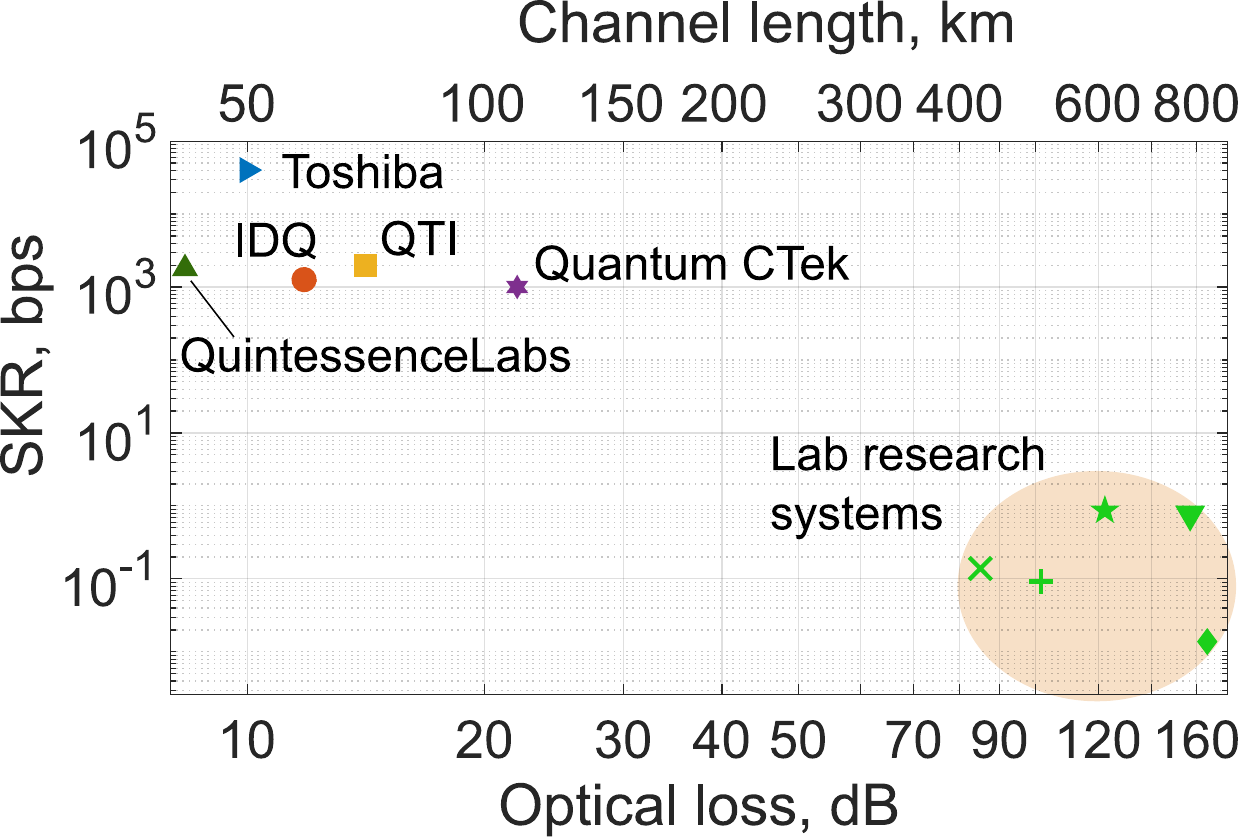}
\centering
\caption{{Secret key rate (SKR) estimates \emph{vs} channel loss/length. Data for QKD vendors is available from their respective websites~\cite{toshiba, idq_qkds, ctek, qti, quislabs}. Data for lab research systems is adapted from Ref.~\cite{wang_twin-field_2022}.}}
\label{fig:skfTypProtocols}
\end{figure}
SKR is useful for comparing the performance of QKD systems or protocols, as also shown in Fig.~\ref{fig:skfTypProtocols} which plots the SKR values as a function of the loss of the quantum channel. The equivalent channel length values assume an optical fiber attenuation coefficient of 0.2 dB/km. In the top-left of the figure are SKR values obtained from online datasheets or brochures of five prominent QKD manufacturers: QuintessenceLabs, Toshiba, ID Quantique (IDQ), Quantum Telecommunications Italy (QTI), and QuantumCTek. The datapoints in the bottom-right of the figure illustrate the reach of state-of-the-art lab-based research systems in various experiments since 2018. We note that typically quoted values from QKD vendors are conservative, i.e., in practice, higher loss may be tolerated without a significant hit to the SKR values. Also, these values are accompanied by an overall failure probability $\epsilon$ which is typically in the $10^{-15}$ to $10^{-9}$ range, which is not always the case with lab research systems. 

Another oft-used parameter or a secret key fraction (SKF), i.e., the number of secret key bits produced per symbol. SKF is more useful for gauging the robustness of QKD implementations to fluctuations, e.g., changes in the channel parameters. For the purpose of data encryption though, SKR is the obvious parameter of interest. 

CVQKD implementations have an advantage over their DV counterparts as they use mostly commercial and off-the-shelf telecommunication equipment and operate at room temperature. However, DVQKD is more loss-tolerant than CVQKD, which means for obtaining reasonable SKF values when the authorized parties are separated by long distances, DVQKD implementations are a preferred choice. 
%%%%%%%%%%%%%%%%%%%%%%%%%%%%%%%%%%%%%%%%%%%%
\subsubsection{Encryption modes}\label{qkd:encmodes}
While OTP encryption offers information-theoretic security, it has several disadvantages when it comes to implementation. The Federal Office for Information Security (BSI) in Germany in fact does not recommend the use of OTP \emph{alone} for encryption~\cite{bsi_RecoKeyLens}. In any case, a correct OTP encryption is feasible only when the SKR matches the data rate. But even the fastest QKD implementation as of today would end up being short by at least two orders of magnitude, if it were to protect a data link operating at a rather basic rate of 100 Gbps. % Even for a rather basic data rate value of 100 Gbps, this would imply a shortfall by at least two orders of magnitude with typical QKD implementations.

Alternatively, one can use the QKD generated key material for implementing other symmetric encryption schemes, such as AES. This also greatly relaxes the requirement on the SKR from the deployed QKD system. For encryption of large volumes of data over a short period of time, the recommendation from official bodies such as National Institute of Standards and Technology (NIST) and BSI vary between a few seconds to a few days~\cite{civiqD2.2, barker_recommendation_2016}. % National Institute of Standards and Technology (USA) recommends refreshing symmetric keys on a daily or weekly basis~\cite{barker_recommendation_2016}. A more recent recommendation from the Federal Office for Information Security (Germany) is to change AES keys after encrypting a Terabit (Tb) of data. 
Specific to AES, typical Ethernet encryption device vendors~\cite{zybersafe_cloak, rohde_n_schwarz} use periods between few seconds to few hours (depending on the data link speed) for renewal of session keys, which were discussed in Section~\ref{bg:prob_key_distb}. Assuming a very short session of just one second, having a pair of AES 256-bit keys (implying a required SKR of at least 512 bps) would enable both encryption of the message-to-send and decryption of the (encrypted) message-just-received. To keep a safety margin, a SKR of $\gtrsim$ 1.0 Kbps from the QKD implementation should suffice for supporting links operating at even Tbps rates. One can observe that most commercial QKD systems visualized in Fig.~\ref{fig:skfTypProtocols} can easily meet this requirement for channel losses until around 10 dB. %In general, QKD can support any typical AES encryption infrastructure quite well. 

It is noteworthy that just like in the case of public key cryptography, quantum computing does provide an advantage in cryptanalysis of AES as well, however, the speedup is merely quadratic~\cite{Grover1996} (compared to exponential~\cite{Shor1994} in case of asymmetric cryptographic algorithms). In other words, quantum computing advances do not compromise the computational security guarantee of AES. 
%Could add something about MACsec here, if needed. 
%%%%%%%%%%%%%%%%%%%%%%%%%%%%%%%%%%%%%%%%%%%%
\subsubsection{Trusted nodes}\label{qkd:trstdnodes}
Just like classical repeaters are utilized in optical telecommunication to regenerate signals that have incurred a large loss after propagating long distances, one could use so-called quantum repeaters to extend the reach of QKD. However, in practice, the current quantum repeater technology is still not sufficiently mature to be integrated with QKD in a beneficial manner. 

Instead, using so-called trusted relays or nodes has gained traction over the last decade~\cite{battelleIDQ, Chen2021short}. In its simplest form, a trusted node links two quantum channels to facilitate key distribution between two authorized parties that otherwise would not have been able to obtain keys due to the high loss (of the combination of the two channels). Several techniques on how to implement AES in such trusted node architectures have been proposed~\cite{civiqD2.2}.
%%%%%%%%%%%%%%%%%%%%%%%%%%%%%%%%%%%%%%%%%%%%
\section{Data centers}
%%%%%%%%%%%%%%%%%%%%%%%%%%%%%%%%%%%%%%%%%%%%
A data center is a facility used to house an organization's IT systems and equipment, including servers, storage systems, network equipment, and other technology infrastructure. Data centers are designed to provide a secure and reliable environment for these systems, and typically include features such as redundant power and cooling systems, fire suppression, and physical security measures.

\subsection{Architectures and classifications}\label{dc:archNclass}
Data centers were conventionally built to support IT functions, including data storage, processing, and management, as well as application and web hosting, cloud computing, and other services. For modern businesses and organizations, data centers offer a widened spectrum of IT services by providing the infrastructure and resources necessary for big data analytics, machine learning, artificial intelligence, and more. They can be operated by the organization that owns the equipment, or by a third-party provider.

\vspace{2ex}
\fbox{
%\colorbox{color2}{
\begin{minipage}{0.92\textwidth}
\textit{\large Data center classification based on ownership and operation}
\begin{itemize}
    \item[$\blacktriangle$] \textbf{On-premises data centers}: The facility and the equipment are located within the organization that owns and operates the data center. 
    \item[$\blacktriangle$] \textbf{Colocation data centers}: The facility is owned and operated by a third-party provider and the organization rents space to house its equipment. 
    \item[$\blacktriangle$] \textbf{Cloud data centers}: The facility and the equipment are owned and operated by a cloud provider and the organization rents access to the infrastructure and services. 
\end{itemize}
\end{minipage}
}
\vspace{2ex}

Also, there are a few common ways that large enterprises typically organize their approach to utilization of the services from data centers: 
\begin{itemize}
    \item \textbf{Centralized}: This approach involves having a single, central data center that serves as the primary location for storing and processing data. This approach is often used by enterprises that need to ensure that their data is stored in a secure location and that it can be easily accessed and managed. 
    \item \textbf{Distributed}: This approach involves having multiple data centers located in different regions. This approach is often used by enterprises that need to ensure that their data is stored in multiple locations for disaster recovery or compliance reasons. They also use data center management software to automate the management of multiple data centers, and to monitor the performance and availability of their data center infrastructure. 
    \item \textbf{Hybrid}: This approach combines elements of both centralized and decentralized data centers, by having a central data center and one or more remote data centers. This approach is often used by enterprises that need to ensure that their data is stored in a secure location and that it can be easily accessed and managed, while also ensuring that their data is stored in multiple locations for disaster recovery or compliance reasons. 
    \item \textbf{Cloud-based}: This approach involves using a cloud provider such as Amazon Web Services (AWS), Microsoft Azure or Google Cloud (GCP) to host and manage data centers, rather than owning and operating them in-house.
\end{itemize}
Based on specific needs and requirements such as ensuring scalability and increasing the level of operational resilience, it is not uncommon to have a mix of these. 
\vspace{-1ex}
\subsection{Connectivity}\label{dc:conn}
Distributed and hybrid data centers are connected through a variety of different technologies, depending on the specific needs and requirements of the enterprise. Some common ways that decentralized data centers are connected include: 
\begin{itemize}
    \item \textbf{Wide Area Network (WAN)}: A WAN is a network that connects multiple locations, such as data centers, over a wide geographic area. WANs are often used to connect decentralized data centers and can be established using a variety of different technologies, such as leased lines, Multiprotocol Label Switching (MPLS), Virtual Private Network (VPN), or satellite links.
    \item \textbf{Software-defined WAN (SD-WAN)}: A SD-WAN is a virtual WAN that connects decentralized data centers using a software-based approach. SD-WANs use multiple connections, such as broadband, cellular, or satellite, to connect data centers and automatically select the best path for data traffic based on real-time network conditions.
    \item \textbf{Cloud Interconnect}: Cloud interconnect enables to securely connect on-premises infrastructure to a cloud provider such as AWS, Azure or GCP. This allows the enterprise to access their data and applications stored in the cloud from their decentralized data centers.
    \item \textbf{Hybrid Cloud Connectivity}: This approach allows for the integration of on-premises infrastructure with public and private clouds, enabling the enterprise to access and manage data across multiple data centers and cloud environments.
    \item \textbf{Direct Peering}: Direct Peering is a way for data centers to connect directly to each other, bypassing the public internet. 
\end{itemize}
The specific technology used to connect data centers will of course depend on the specific needs and requirements of the enterprise, such as the amount of data that needs to be transferred, the level of security required, and the available bandwidth.
\vspace{-1ex}
%%%%%%%%%%%%%%%%%%%%%%%%%%%%%%
\subsection{Certification}
%%%%%%%%%%%%%%%%%%%%%%%%%%%%%%
Like for most industries, data center certification can be a useful tool that adds credibility on aspects ranging from operations and energy efficiency to security and reliability. There are several organizations that certify data centers to ensure that they meet certain standards and guidelines for infrastructure, redundancy, and availability. 

\vspace{2ex}
\fbox{
%\colorbox{color2}{
\begin{minipage}{0.92\textwidth}
\textit{\large Data center certification/standards organizations across the world}
\begin{itemize}
    \item[$\blacktriangle$] \textbf{Telecommunications Industry Association (TIA)}: This organization provides standards and guidelines for data centers, including the TIA-942 standard for data center infrastructure. 
    \item[$\blacktriangle$] \textbf{American National Standards Institute (ANSI)}: This organization provides standards and guidelines for data centers, including the ANSI/BICSI 002-2014 standard for data center design and infrastructure.
    \item[$\blacktriangle$] \textbf{International Organization for Standardization (ISO)}: This organization provides international standards for data center infrastructure, including the ISO/IEC 27001 standard for information security management.
    \item[$\blacktriangle$] \textbf{European Telecommunications Standards Institute (ETSI)}: This organization provides standards and guidelines for data centers in Europe, including the EN 50600 series of standards for data center infrastructure.
\end{itemize}
\end{minipage}
}
%%%%%%%%%%%%%%%%%%%%%%%%%%%%%%%%%%%%
\subsubsection{The Uptime Institute}
%%%%%%%%%%%%%%%%%%%%%%%%%%%%%%%%%%%%
The Uptime Institute (\url{https://uptimeinstitute.com/}) is an unbiased advisory organization that provides certification for data centers based on a tiering model. They also provide consulting services to help data centers improve their infrastructure and achieve higher tier certifications. The certification from Uptime is currently the most widely used and their tiering model consists of 4 tiers:
\begin{itemize}
    \item Tier I: This is the most basic level of data center infrastructure. It typically includes a single path for power and cooling, and little or no redundancy. Downtime is not uncommon, and recovery can take an extended period of time.
    \item Tier II: This level of data center infrastructure includes redundant components, such as multiple power and cooling systems. Downtime is less common than Tier I, but recovery may still take an extended period of time.
    \item Tier III: This level of data center infrastructure includes multiple paths for power and cooling, and redundancy for critical systems. Downtime is extremely rare, and recovery can be completed in a relatively short period of time.
    \item Tier IV: This is the highest level of data center infrastructure, and it is characterized by multiple paths for power and cooling, and redundancy for all systems. Downtime is extremely rare, and recovery can be completed in a very short period of time.
\end{itemize}
The Uptime Institute also has a provision for a concurrent maintainability Tier III and Tier IV, i.e., the maintenance of and upgrades in these tiers can be done without interrupting the data center services.
%\vspace{-2ex}
%%%%%%%%%%%%%%%%%%%%%%%%%%%%%%%%
\section{QKD integration in data centers} 
%%%%%%%%%%%%%%%%%%%%%%%%%%%%%%%%
Since photons are currently the only reliable carriers of quantum information over long distances, it is obvious that in order to integrate QKD as a security service, the communication nodes should have access to an optical channel, for instance, a fiber-optic cable or a line-of-sight link. Furthermore, the sensitivity of the quantum signal to optical loss and noise implies the channels should minimize lossy connections, and definitely be free of active components such as amplifiers. 
%Several QKD demonstrations (with some of the recent ones mentioned in section~\ref{sec:mot}) over the last two decades have firmly established that QKD---under the general constraints mentioned above---is a mature technology for secure communications. 
%%% perhaps something about costs ?? 

Taking these constraints into account, a growing number of demonstrations in the last few years have showcased applications of QKD for secure communications in real life situations. Notably, in the context of fibers, several successful studies have proven co-existence of quantum signals with classical traffic for both DV and CV systems~\cite{dynes_cambridge_2019, Eriksson2019, Wang_QAN_2021}, implying that dark fibers are no longer necessary for QKD. 

The pertinent question then is: Does QKD offer anything useful to the data center security infrastructure and if so, how? On the one hand, investing in any typical emerging technology always carries some risks, and particularly for QKD, it is entirely possible that the eventual rewards are not very high while the costs are. Yet, on the other hand, as elucidated in the first half of this white paper, the issue of data security cannot be overlooked anymore. In that regard, reputational damage due to data breaches or leaks would be as significant as the legal and economic consequences. With the growing competition for hosting services, a higher level of security thus may be a differentiator. More precisely, the motivation would be to lower the risks of a security breach in a future where quantum computers become a reality. In fact, QKD has quite a natural space in such a crypto-agile environment. 

A hidden advantage is that the step of QKD integration shall trigger a risk assessment that obligates the data center to review their current infrastructure. Even if the eventual outcome is to not adopt QKD, an exploration of new tools and opportunities to enhance the security can nevertheless be expected. 
\vspace{-1ex}
%%%%%%%%%%%%%%%%%%%%%%%%%%%%%%%%%%%%%
\subsection{Market potential and scalability}
%%%%%%%%%%%%%%%%%%%%%%%%%%%%%%%%%%%%%
% Add sentences related to QKD from the datacenter.docx from Jesper 
% While QKD can grant the strongest form of data confidentiality (under the assumptions that the laws of physics are correct and the implementation of the hardware matches the theoretical model), the practical costs and challenges of deploying such technology are also fairly considerable. 
Typical customers purchasing QKD systems are those with high value data that require security over long periods, e.g., when the lifetime of data is enforced by a regulatory entity, as also illustrated in Fig.~\ref{fig:data_ret}. %"Following a dialogue with an expert of a EU quantum ecosystem", 
Colocation or cloud data centers (see section~\ref{dc:archNclass}) with a client base in this category would benefit from offering QKD as a service as an additional protection layer. In the case of an on-premises data center, it could well be that there is a desire to implement the strong security offered by QKD as a default. In that sense, there are two ways for early adopters to realize the market potential: 
\begin{enumerate}
    \item Targeting dedicated users with real security needs and interest in applying emerging technology as an additional layer, i.e., implement QKD as a service for each single use case.
    \item Create proposition for an easy plug-and-play QKD-as-a-service so that quantum keys can be queried if and when needed.
\end{enumerate}
Note that the two options can also be part of a development plan, i.e., after a successful proof-of-concept trial, the data center can provide the tools to create the proposition based on the learning.

The step towards scalability is dictated by standards and regulation, which have a long-term role of increasing security and lowering liability. Nonetheless, implementing QKD today will not increase the client's security from a regulatory perspective as there are no standards backing the implementation. This reduces the incentive in upgrading the security systems: The resultant obsolete security systems may not affect liability but do not grant a high level of security. This is however a general problem even with the post-quantum cryptographic algorithms, slated to replace the current asymmetric algorithms (due to their vulnerability to quantum computers). As these are also not sufficiently mature, one should strongly consider scaling the adoption of QKD to keep all options open and ensuring crypto-agility. 
%One last fundamental step towards scalability is dictated by standards. The role of standards and regulation is to grant security and lower liability, therefore implementing QKD today will not increase the client security from a regulatory prospective. This is because there are no standards backing the implementation and no incentive in upgrading the security system. This results in sometimes obsolete security systems that lower the liability, but do not grant a high level of security. This is a central problem that is being addressed (e.g. NSA recommendations [reference] or NIST standards for PQC [reference]), but it is not mature enough and should be strongly considered for scaling the adoption of QKD.
%\vspace{-1ex}
\subsection{Practical feasibility}
With regards to integration in data centers, the distributed and hybrid approaches on how a large enterprise may avail data center services (mentioned in section~\ref{dc:archNclass}) are the most interesting for QKD. Even more specifically, the direct peering connectivity solution listed in section~\ref{dc:conn}, which allows for faster and more secure data transfer between data centers, is possibly the most natural candidate where QKD can add value. The point-to-point link in Fig.~\ref{fig:scheme} conveys a possible implementation methodology.

In fact, there have already been demonstrations in Israel, Austria, and Singapore~\cite{quantLRnvidia, Zatoukal2021, OpenGov-NQSN} in which QKD was incorporated in a distributed and a cloud-based environment.  Below we present the details of two of these use cases. We note first that as discussed in section~\ref{qkd:encmodes} and illustrated in Fig.~\ref{fig:skfTypProtocols}, data center connections with optical losses around 10 dB can easily be served by commercially available QKD systems. While one can soon expect a new range of QKD devices that can tolerate higher losses (blob with green markers in the bottom right of Fig.~\ref{fig:skfTypProtocols}), an alternative that is currently feasible is the trusted node architecture, elaborated in section~\ref{qkd:trstdnodes}. 
%\vspace{-2ex}
%%%%%%%%%%%%%%%%%
\subsubsection{Use case I: National Quantum-Safe Network (Singapore)}
%%%%%%%%%%%%%%%%%
At the Q2B conference in Dec. 2022, AWS and Horizon Quantum Computing (Horizon) jointly announced the deployment of the first QKD link in the Singapore National Quantum-Safe Network (NQSN). 

\begin{figure}[!thbp]
\centering
\includegraphics[width=0.95\linewidth]{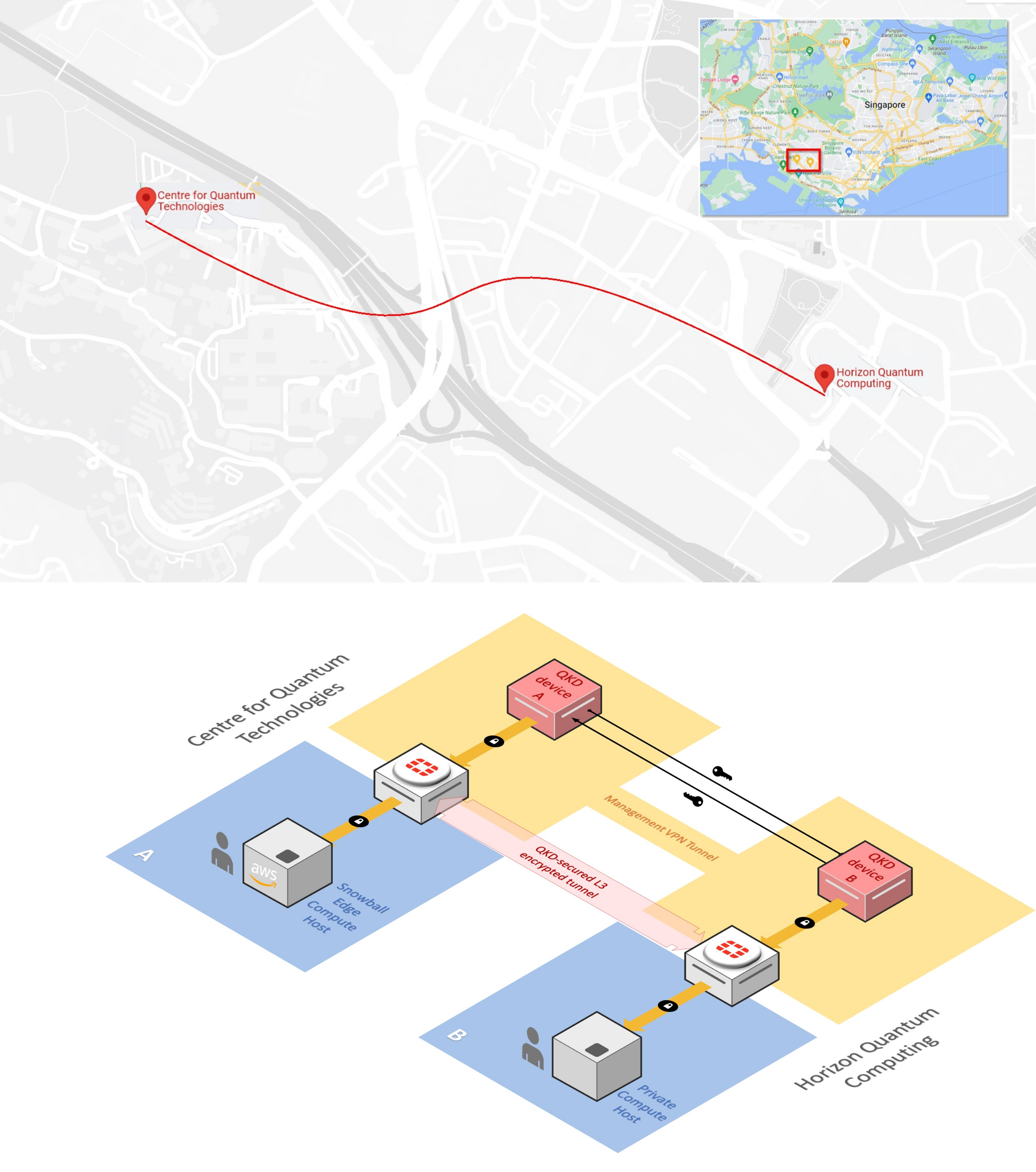}
\caption{Schematic overview of the first deployed QKD link of the Singapore National Quantum-Safe Network. At the CQT side, the computing endpoint is an Amazon Elastic Compute Cloud (EC2) instance encapsulated inside an AWS Snowball Edge Compute Optimized device. AWS Hybrid-Edge services extend AWS infrastructure and services into the edge, helping to run and securely operate applications in locations that lack consistent network connectivity to AWS. At the other end of the connection, Horizon uses an on-premises Intel-based server as an endpoint. Both sides communicate securely via an IPsec tunnel, with the endpoints consuming the locally produced QKD keys. This effectively sets a quantum-safe link that Horizon will be using to build other use cases. Figure courtesy of Juan Moreno, AWS Center for Quantum Networking.}
\label{fig:AWS-usecase}
\end{figure}

\vspace{0.5ex}
\colorbox{color2}{
\begin{minipage}[t]{0.43\textwidth}
``{\small At the AWS Center for Quantum Networking, we aim to explore quantum communication technologies in the context of AWS services. These technologies can enable quantum enhanced computing, communication, and sensing with capacities uniquely accessible using quantum techniques. We plan to test, develop and deploy this technology in a manner that brings value to AWS customers while exploring new frontiers of technology.}'' 

\textit{\mbox{Juan Moreno}, }

\textit{AWS Center for Quantum Networking}
\end{minipage}
}
\hspace{0.06\textwidth}
\begin{minipage}[t]{0.44\textwidth}
Per communications about the project~\cite{OpenGov-NQSN, aws_blog_post}, the NQSN aims to enhance network security for critical infrastructures with quantum technology while also serving as a robust platform for public-private collaboration. The deployment plan is to have 10 fiber connected network nodes installed across Singapore, including university, government, and private company premises. In the first deployed link of the network, schematically illustrated in Fig.~\ref{fig:AWS-usecase}, QKD is used to distribute keys over a quantum-
\end{minipage}
\vspace{0.2ex}

\noindent safe link between an AWS Snowball Edge compute host located at the Centre for Quantum Technologies (CQT) and an on-premises compute host at Horizon. The proof-of-concept demonstrates a point-to-point VPN established over this link, evaluating the technological status of current QKD systems and their fit for cloud infrastructure. 
% AWS' participation in the PoC is through the AWS Center for Quantum Networking, established in 2022. Juan Moreno of AWS CQN says, \textit{“AWS created the Center for Quantum Networking to explore usecases and do research on what is possible. We want to make sure that our customers are protected from a potential quantum attack, whenever that becomes a reality (we believe it is not a matter of “if” but “when”). QKD is one of the options to ensure that protection.”}

\paragraph{Technical details}
Two mirrored network stacks are setup, divided into management and service segments via a FortiGate 100F Next Gen Firewall, that has the enhanced capability to create an encrypted IPsec tunnel capable of consuming quantum-generated encryption keys. On the management network, QKD device pair is used for secure key exchange over fibers with a maximum loss of 12 dB (typically up to 50 km), making it a good fit for metropolitan coverage of an area of the size of Singapore (50 km east to west). This device also integrates a Key Management System that handles key requests and key transfers between QKD optical systems and the FortiGate Next Gen Firewall. The service network terminates at the aforementioned paired compute nodes.
%%%%%%%%%%%%%%%
\subsubsection{Use case II: Medical Data Protection (Austria)}\label{qkd_dc:graz}
%%%%%%%%%%%%%%%
In the context of data center protection, an interesting demonstration was conducted in Austria in 2020 within the framework of the European Union project `OpenQKD'~\cite{openqkd}. With an ambition of enabling secure collaboration on patient data between remote hospitals to improve diagnosis and treatment, the use case ``Medical Data Protection Graz'' demonstrated sharing of patient-sensitive medical data between two hospitals with information theoretic security.

\begin{figure}[!hbpt]
\centering
\includegraphics[width=0.97\linewidth]{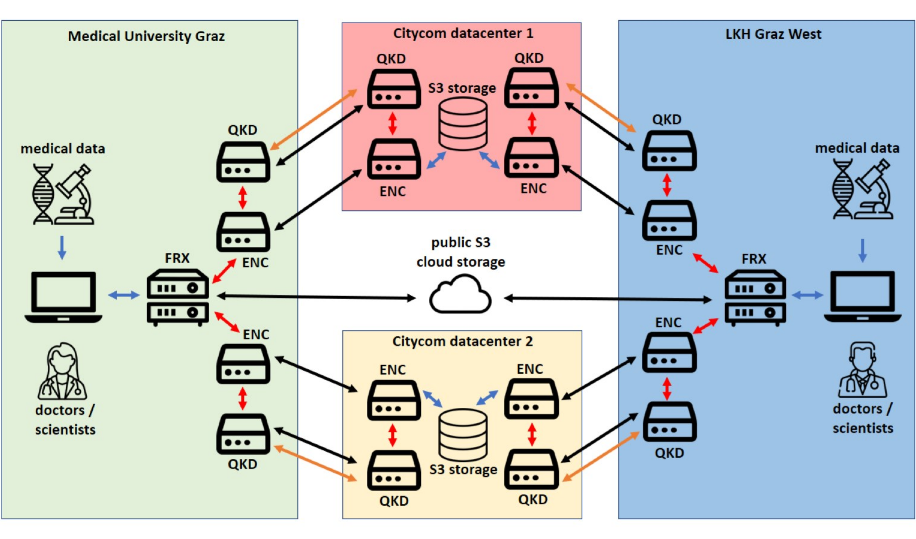}
\caption{Overview illustration of the technical implementation of the Medical Data Protection Graz use case. Source: \protect\url{https://openqkd.eu/wp-content/uploads/2020/01/UC21.png}}
\label{fig:Graz-usecase}
\end{figure}

In the use case, schematically illustrated in Fig.~\ref{fig:Graz-usecase}, two hospitals securely shared medical data using a combination of data fragmentation and quantum key distribution. 
The data was split in three fragments whereof two were transmitted securely to data centers using QKD while the third fragment was transmitted to a traditional https-protected storage, simulating S3 public cloud storage. The distributed storage, enabled by data fragmentation, provided \emph{security for data at rest} at the three storage locations and resilience in case of unavailability of one of the storage servers. To elaborate, access to any two of the three locations would have been sufficient to reconstruct the data. In addition, the use of QKD for symmetric encryption provided \emph{security for data in transit} between end users at the hospitals and the in-between data centers.

\paragraph{Technical details}
Data in transit was encrypted using 10\,Gbit high-speed encryptors from ADVA, taking as input keys provided by QKD systems from Toshiba and ID Quantique, and the length of the fiber links connecting end user and storage sites varied between 9\,km and 20\,km with losses of 3.3\,dB to 8.1\,dB. The QKD key rates ranged between 1\,kbit/s and 2.2\,Mbit/s, however, in continuous operation mode the encryptor demand was only 256 bits per 10 minutes. The block-size of data encrypted with the same AES-key was 6\,Tbit. For secret sharing of stored data, fragmentiX appliances were employed. For further information, see~\cite{Zatoukal2021}.

% risk assessment, policy revision : obligated to look at your current infrastructure, looking into new tools. QKD is a differentiator, an advertising tool, crypto agility: adapting to threat quickly 

\section{Conclusion}
While QKD can grant the strongest form of data confidentiality possible as of today, the practical costs and challenges of deploying the technology are also fairly considerable. However, both the scientific community and industry is working towards lowering costs, improving the performance, and engaging in standardization, so deploying QKD in the near future can be expected to become more affordable and useful. %(under the assumptions that the laws of physics are correct and the implementation of the hardware matches the theoretical model)

In the context of securing the flow of information for data centers, we have described how QKD systems can support AES encryption for data protection on high-capacity links. For data center owners and operators, who are planning to investigate the adoption of QKD for data security, we believe that QKD can take a role somewhere between being an advertising tool to an actual differentiator that ensures crypto-agility. In either case, an exploration of this technology is a strongly recommended activity, as measures to counter the quantum computing threat might eventually be quite limited. %and understanding how to implement it in the current infrastructure

\section*{Acknowledgements}
This study was supported by Copenhagen Fintech as part of the ``National Position of Strength programme for Finans \& Fintech'', funded by the Danish Ministry of Higher Education and Science. Funding within the QuantERA II Programme (European Union’s Horizon 2020 research and innovation programme under Grant Agreement No 101017733) CVStar is also acknowledged. The authors additionally acknowledge support from Innovation Fund Denmark through the project CryptQ (0175-00018B). 

\newpage
\bibliographystyle{unsrt}
\bibliography{library}

\end{document}